\documentclass[fleqn,10pt]{wlscirep}
\usepackage[utf8]{inputenc}
\usepackage[T1]{fontenc}
\usepackage{graphicx}
\usepackage{xcolor}
\usepackage{subcaption}
\usepackage{caption}
\usepackage{listings}
\title{Magnum.np.distributed: Accelerating Finite Diﬀerence Micromagnetic Simulations with Multiple GPUs}

\author[1,*]{Tsz Chung Cheng}
\author[1]{Yuichiro Kurokawa}
\author[1]{Hiromi Yuasa}
\affil[1]{Kyushu University, Graduate School of Information Science and Electrical Engineering, Fukuoka, 819-0382, Japan}
\affil[*]{jed.cheng@mag.ed.kyushu-u.ac.jp}

\lstdefinestyle{magnumpython}{
    language=Python,
    basicstyle=\ttfamily\footnotesize,
    keywordstyle=\color{blue!80!black}\bfseries,
    stringstyle=\color{red!60!black},
    commentstyle=\color{gray}\itshape,
    numberstyle=\tiny\color{gray},
    numbers=left,
    numbersep=6pt,
    stepnumber=1,
    firstnumber=1,
    showstringspaces=false,
    breaklines=true,
    tabsize=4,
    frame=single,
    framerule=0.4pt,
    rulecolor=\color{gray!40},
    backgroundcolor=\color{gray!6},
    xleftmargin=1.5em,
    framexleftmargin=1.5em,
    morekeywords={True,False,None},
}

\begin{abstract}
Micromagnetic simulations are essential tools in nanomagnetism and spintronics research. Although widely adopted solvers like Mumax3 and the Python-native magnum.np use GPU acceleration to improve performance, these tools are limited to single-device computation. In this work, we present the first Python-native multi-GPU micromagnetic framework by extending magnum.np with PyTorch Distributed. This leverages high-speed communication and computation across multiple GPUs while retaining the benefits of ease of installation, platform-agnostic design, and compatibility with Python. For computationally intensive demagnetisation effective-field calculations, we achieve a 7.0x speedup across 8 GPUs connected via NVLink, whereas Halo exchange required for Heisenberg exchange shows limited scaling due to kernel dispatch latency. We also demonstrated the framework's versatility by achieving a 6.8x speedup in demagnetisation field computation on CPU with NUMA pinning via the MPI backend of PyTorch Distributed. Faster turnaround times will enable researchers to explore larger, more complex systems and accelerate the design cycle for novel spintronic devices. Source code: \url{https://gitlab.com/jedcheng/magnum-np-distributed}
\end{abstract}
\begin{document}

\flushbottom
\maketitle

\thispagestyle{empty}

\section*{Introduction}
Micromagnetic simulations are widely used in nanomagnetism and spintronics research. \cite{leliaert2019tomorrow} Researchers rely on finite-difference method (FDM) simulations to compare experimental results with theoretical understanding \cite{hassan2024dipolar, garlow2019quantification} and to design neuromorphic computing systems. \cite{dion2024ultrastrong, song2020skyrmion} Mumax \cite{vansteenkiste2011mumax} and the subsequent Mumax3 \cite{Vansteenkiste2014} have been the most widely used open-source micromagnetic simulation software that leverages graphical processing units (GPUs) to accelerate computation compared to existing central processing unit (CPU)- based solvers, such as OOMMF. \cite{donahue1999oommf} Mumax3 uses Go, and OOMMF uses TCL for scripting, which poses a significant challenge for researchers who need to interchange simulation scripts between platforms and introduces overhead when analysing simulation data with Python. Ubermag provides a unified Python interface for interacting with the underlying Mumax3 and OOMMF simulations and bundles various analysis tools. \cite{beg2022} Mumax+ built on the success of Mumax3 and recognised the workflow problem by wrapping Python modules around the C++ and CUDA-based simulation core, enabling flexible runtime analysis with a wide range of Python packages. \cite{moreels2026mumax+}\\

Magnum.np is a Python-native FDM solver that uses PyTorch for high-performance computations without vendor lock-in in CUDA-based software.\cite{bruckner2023magnum, paszke2019pytorch} The just-in-time (JIT) compiler uses TorchDynamo to trace the computational graph and TorchInductor to automatically compile machine code for different hardware at runtime. \cite{torch_compile} This allows new features to be implemented easily without writing or compiling C++ code. The Python-native design also enabled Python-based analysis during simulation, which is great for designing neuromorphic computing devices. PyTorch also supports autograd for gradient computation, enabling the simulator to solve inverse problems to optimise for certain design goals. The most beneficial design feature in our experience is the ease of switching between CPU and GPU, or between single- and double-precision, all on a single line of code. NeuralMag takes the same idea to JAX and optimises the framework for solving inverse problems. \cite{abert2025neuralmag}\\

However, all the FDM simulations mentioned only use one device for computation. Boris is a C++ and CUDA-based simulator that supports multiphysics and atomistic simulation. \cite{lepadatu2020boris} The simulator was later upgraded to leverage multiple NVIDIA GPUs for further speedup. \cite{lepadatu2023accelerating} However, the requirement to compile the CUDA code from source on Linux systems created a major barrier to practical use in high-performance computing (HPC) systems. In this work, we extended magnum.np with PyTorch distributed to combine the benefits of high-speed computation from multiple GPUs while preserving the flexibility and ease of installation of the Python-native package. PyTorch Distributed provides production-grade collective and point-to-point communication via multiple backends with support for PyTorch operations such as JIT compilation and the autograd feature. Currently, we are targeting the NCCL backends for NVIDIA GPUs connecting via NVLink and InfiniBand and the MPI backend for CPU-based compute, while remaining platform-agnostic by avoiding CUDA-based optimisations. The former can be installed via pip without any compilation. Although the latter requires compiling PyTorch from source, the PyTorch team conducts extensive testing to eliminate potential compilation issues.

\section*{Implementation}
Micromagnetic simulations model the magnetic dynamics via the Landau-Lifshitz-Gilbert equation:
 \begin{equation}                                                                                                                                                                            
      \frac{d\mathbf{M}}{dt} = -\frac{\gamma}{1+\alpha^2}                                                                                                                                     
      \left[\mathbf{M} \times \mathbf{H}_\mathrm{eff}                                                                                                                                         
      + \alpha\, \mathbf{M} \times \left(\mathbf{M} \times \mathbf{H}_\mathrm{eff}\right)\right]                                                                                              
  \end{equation} 
The time derivative of the magnetization vector M is given by the gyromagnetic ratio $\gamma$, the Gilbert damping constant $\alpha$, and the sum of various physical phenomena acting as the effective field Heff. \cite{landau1935theory, gilbert2004phenomenological}\\

In the distributed implementation, the mesh (including the M and Heff) is sharded into contiguous slabs along the x-axis with one slab per rank. In this setup, most of the effective fields act only on local data or are perpendicular, such as the Slonczewski spin transfer torque \cite{abert2019micromagnetics, slonczewski1996current} and uniaxial anisotropy. They do not require magnetisation of neighbouring cells and can therefore be applied directly to distributed simulations. However, the Heisenberg exchange field requires the knowledge of the neighbouring magnetisation as given by the saturation magnetisation $M_\mathrm{sat}$, permeability of free space $\mu_0$ and the exchange stiffness $A_{ex}$: \cite{abert2019micromagnetics}
  \begin{equation}                                                                                                                                                                            
      \mathbf{H}_\mathrm{eff,exchange} = \frac{2}{\mu_0 M_\mathrm{sat}}
      \nabla \cdot \left(A \nabla \mathbf{M}\right)                                                                                                                                           
  \end{equation}

The same applies to the Dzyaloshinskii-Moriya Interaction (DMI) \cite{moriya1960anisotropic, dzyaloshinsky1958thermodynamic} effective field as given by: \cite{abert2019micromagnetics}

  \begin{equation}                                                                                                                                                                            
      \mathbf{H}_\mathrm{eff,DMI} = \frac{2D}{\mu_0 M_\mathrm{sat}}
      \sum_{k=x,y,z} \mathbf{e}_k^\mathrm{DMI} \times \frac{\partial \mathbf{M}}{\partial k}
  \end{equation}   

With the DMI constant D and DMI vector $e_k^{DMI}=(e_y,-e_x,0)$ for interfacial DMI or $(e_x,e_y,e_z)$ for bulk DMI.\\

Distributed versions of the two effective fields were created, with each slab having two extra cells at each end of the x-direction as a halo region. Communication is done via a batched, non-blocking, point-to-point communication. Unfortunately, point-to-point communications are not optimised by TorchDynamo. The halo communication must be placed outside the JIT compilation, leaving only the computation to be compiled.\\

\begin{figure}[ht]
	\centering
	\includegraphics[width=0.99\textwidth]{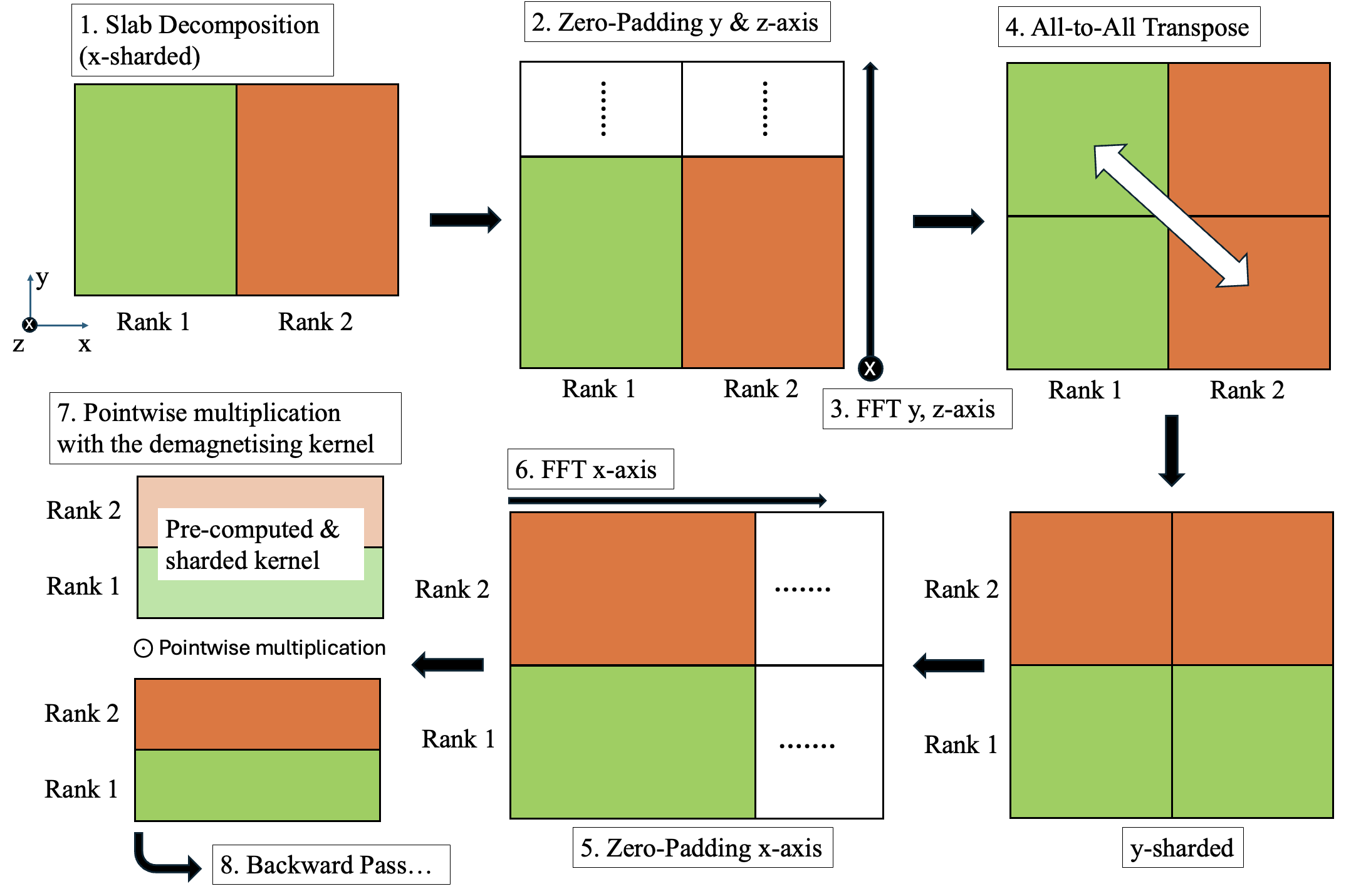}
	\caption{Schematic diagram of the distributed 3D FFT implementation. The magnetisation is first split along the x-axis using slab decomposition. The magnetisation is padded, and the FFT is computed along the y and x axes. An all-to-all transpose is then performed to shard the vectors along the y-axis, followed by zero-padding and FFT computation along the x-axis. Then, a point-to-point multiplication is performed using the demagnetisation kernel. For the backward pass, the steps are reverted with truncation replacing padding and inverse FFT replacing the forward FFT.}
	\label{fig:fig1}
\end{figure}

The demagnetization field is the most computationally intensive and a global interaction. The effective field calculation is done via a discrete convolution of a kernel N computed using Newell’s formula as inherent from the non-distributed magnum.np: \cite{bruckner2023magnum, abert2019micromagnetics, newell1993generalization, abert2015full}
  \begin{equation}                                                                                                                                                                            
      \mathbf{H}_\mathrm{eff,demag}(\mathbf{x}) =                                                                                                                                             
      \int d\mathbf{x}'\; \mathbf{N}(\mathbf{x} - \mathbf{x}')\, \mathbf{M}(\mathbf{x}')
  \end{equation}   
The 3D FFT and inverse FFT passes are computationally and communication-intensive, and researchers have been exploring algorithms to accelerate them using multiple devices. \cite{dalcin2019fast, ayala2022performance} As illustrated in Figure~\ref{fig:fig1}, in our implementation, because the distributed mesh is sharded along the x-axis, zero-padding and 2D FFT were first performed in the y- and z-directions. After an all-to-all transpose, padding, and an FFT in the x-direction, a pointwise multiplication with the now y-directionally sharded demagnetisation kernel is computed. The inverse pass reverses the step by first computing the 1D inverse FFT along the x-direction. The output is then truncated for another all-to-all transpose, followed by a 2D inverse FFT along the y and z directions. Boris provides an option to carry out the all-to-all transpose with halved precision (i.e., compute in double (single) precision and communicate in single (half) precision). However, our testing showed degraded performance with the additional steps of halving the precision and normalising the values by dividing by the saturation magnetisation. The feature to halve precision before transfer was thus removed to maintain code readability. It is attributed to the JIT's current inability to process complex numbers, which prevents it from fusing multiple GPU kernel calls into a single call.\\

To advance the LLG equation, the original non-distributed magnum.np was designed to support multiple differential equation solvers, such as those in the SciPy and TorchDiffEq packages. But only the Runge-Kutta-Fehlberg Method (RKF45) is supported for now. The distributed implementation globally reduces the maximum error before computing the next step size, ensuring that all ranks advance by the same time step. The Barzikai-Borwein algorithm for minimising the system energy also underwent similar changes so that all ranks terminate together after the convergence criteria are met. \cite{ayala2022performance} \\

The distributed implementation is also designed to reduce the need for code changes, but changes are still necessary. Here is the comparison between the simulation script of the original and the distributed version of the muMag standard problem 4 \cite{sp4}:

% ============================================================
% Block 1 — Additional imports
% ============================================================
\noindent\textbf{Additional imports required for the distributed version:}

\vspace{0.5em}
\noindent
\begin{minipage}[t]{0.50\textwidth}
\lstset{style=magnumpython, firstnumber=1}
\begin{lstlisting}
from magnumnp import *
from magnumnp.solvers.distributed_llg \
    import DistributedLLGSolver
from magnumnp.solvers.distributed_minimize \
    import DistributedMinimizerBB
import torch
import torch.distributed as dist
\end{lstlisting}
\end{minipage}%
\hfill
\begin{minipage}[t]{0.47\textwidth}
\lstset{style=magnumpython, firstnumber=1}
\begin{lstlisting}
from magnumnp import *
import torch
\end{lstlisting}
\end{minipage}

\vspace{0.4em}
\noindent\textbf{The process group must be initialised before constructing any distributed object:}
\noindent 
\vspace{0.2em}

\noindent
\begin{minipage}[t]{0.50\textwidth}
\lstset{style=magnumpython, firstnumber=8}
\begin{lstlisting}
dist.init_process_group(backend="nccl")
\end{lstlisting}
\end{minipage}%
\hfill
\begin{minipage}[t]{0.47\textwidth}
\phantom{x}% no equivalent in serial script
\end{minipage}

% ============================================================
% Block 2 — Mesh, state, and material
% ============================================================
\vspace{0.8em}
\noindent\textbf{Replace \texttt{Mesh} and \texttt{State} with their distributed counterparts:}

\vspace{0.5em}
\noindent
\begin{minipage}[t]{0.50\textwidth}
\lstset{style=magnumpython, firstnumber=10}
\begin{lstlisting}
dt = 1e-11
n  = (100, 25, 1)
dx = (5e-9, 5e-9, 3e-9)
dmesh = DistributedMesh(n, dx,
            backend="nccl")
state = DistributedState(dmesh)

state.material = {
    "Ms":    8e5,
    "A":     1.3e-11,
    "alpha": 0.02,
}
\end{lstlisting}
\end{minipage}%
\hfill
\begin{minipage}[t]{0.47\textwidth}
\lstset{style=magnumpython, firstnumber=10}
\begin{lstlisting}
dt = 1e-11
n  = (100, 25, 1)
dx = (5e-9, 5e-9, 3e-9)
mesh  = Mesh(n, dx)
state = State(mesh)

state.material = {
    "Ms":    8e5,
    "A":     1.3e-11,
    "alpha": 0.02,
}
\end{lstlisting}
\end{minipage}

% ============================================================
% Block 3 — Magnetisation initialisation
% ============================================================
\vspace{0.8em}
\noindent\textbf{Magnetisation initialisation via a helper function broadcast from rank~0:}

\vspace{0.5em}
\noindent
\begin{minipage}[t]{0.50\textwidth}
\lstset{style=magnumpython, firstnumber=22}
\begin{lstlisting}
def init_m(global_shape):
    m = torch.zeros(global_shape + (3,))
    m[1:-1, :, :, 0]    = 1.0
    m[(-1, 0), :, :, 1] = 1.0
    return m

state.init_m_from_rank0(init_m)
\end{lstlisting}
\end{minipage}%
\hfill
\begin{minipage}[t]{0.47\textwidth}
\lstset{style=magnumpython, firstnumber=22}
\begin{lstlisting}
state.m = state.Constant([0,0,0])
state.m[1:-1,:,:,0]    = 1.0
state.m[(-1,0),:,:,1]  = 1.0
\end{lstlisting}
\end{minipage}

% ============================================================
% Block 4 — Field terms, minimiser, solver, and time loop
% ============================================================
\vspace{0.8em}
\noindent\textbf{Field terms, minimiser, solver, and time-integration loop:}

\vspace{0.5em}
\noindent
\begin{minipage}[t]{0.50\textwidth}
\lstset{style=magnumpython, firstnumber=30}
\begin{lstlisting}
demag    = DistributedDemagField()
exchange = DistributedExchangeField()
external = ExternalField(
    [-24.6e-3/constants.mu_0,
     +4.3e-3/constants.mu_0, 0.0])

minimizer = DistributedMinimizerBB(
    [demag, exchange])
minimizer.minimize(state)

llg = DistributedLLGSolver(
    [demag, exchange, external])
logger = DistributedScalarLogger(
    "data/log_dist.dat", ["t", "m"])

for i in torch.arange(0, 1e-9, dt):
    llg.step(state, dt)
    logger << state  # all ranks

dist.destroy_process_group()
\end{lstlisting}
\end{minipage}%
\hfill
\begin{minipage}[t]{0.47\textwidth}
\lstset{style=magnumpython, firstnumber=30}
\begin{lstlisting}
demag    = DemagField()
exchange = ExchangeField()
external = ExternalField(
    [-24.6e-3/constants.mu_0,
     +4.3e-3/constants.mu_0, 0.0])

minimizer = MinimizerBB(
    [demag, exchange])
minimizer.minimize(state)

llg = LLGSolver(
    [demag, exchange, external])
logger = Logger(
    "data", ["t", "m"])

for i in torch.arange(0, 1e-9, dt):
    llg.step(state, dt)
    logger << state
\end{lstlisting}
\end{minipage}

\vspace{0.8em}
\noindent The distributed script is launched via:

\begin{lstlisting}[style=magnumpython, language=bash,
    morekeywords={torchrun,mpirun},
    keywordstyle=\color{teal}\bfseries]
# NCCL backend (GPU)
torchrun --nproc_per_node=8 run_dist.py --backend=nccl

# MPI backend (CPU, NUMA-pinned)
mpirun -genv I_MPI_PIN_DOMAIN=numa -n 8 python run_dist.py --backend=mpi
\end{lstlisting}

\section*{Results}
\subsection*{Validation}
The muMAG standard problem 4 ensures the demagnetisation and exchange effective field and the LLG dynamics via a magnetic reversal in a permalloy magnet. \cite{sp4} We performed distributed simulations with 2, 4 and 8 ranks in double precision and compared them to the original magnum.np implementation, as well as Mumax3 (single-precision only). The comparison over 1 ns of simulation is displayed in Figure 2. The agreement verifies the distributed demagnetisation and exchange field, the energy minimiser, and the distributed LLG-RKF45 solver implementation. The standard problems for DMI \cite{cortes2018proposal} and domain wall pinning \cite{heistracher2022proposal} were also used to verify the Halo exchange for DMI and overall implementation of inhomogeneous material parameters, respectively. The data and scripts to reproduce the results are available in the project repository, along with additional comparisons between the original non-distributed magnum.np and the distributed implementations.

\begin{figure}[ht]
	\centering
	\includegraphics[width=0.99\textwidth]{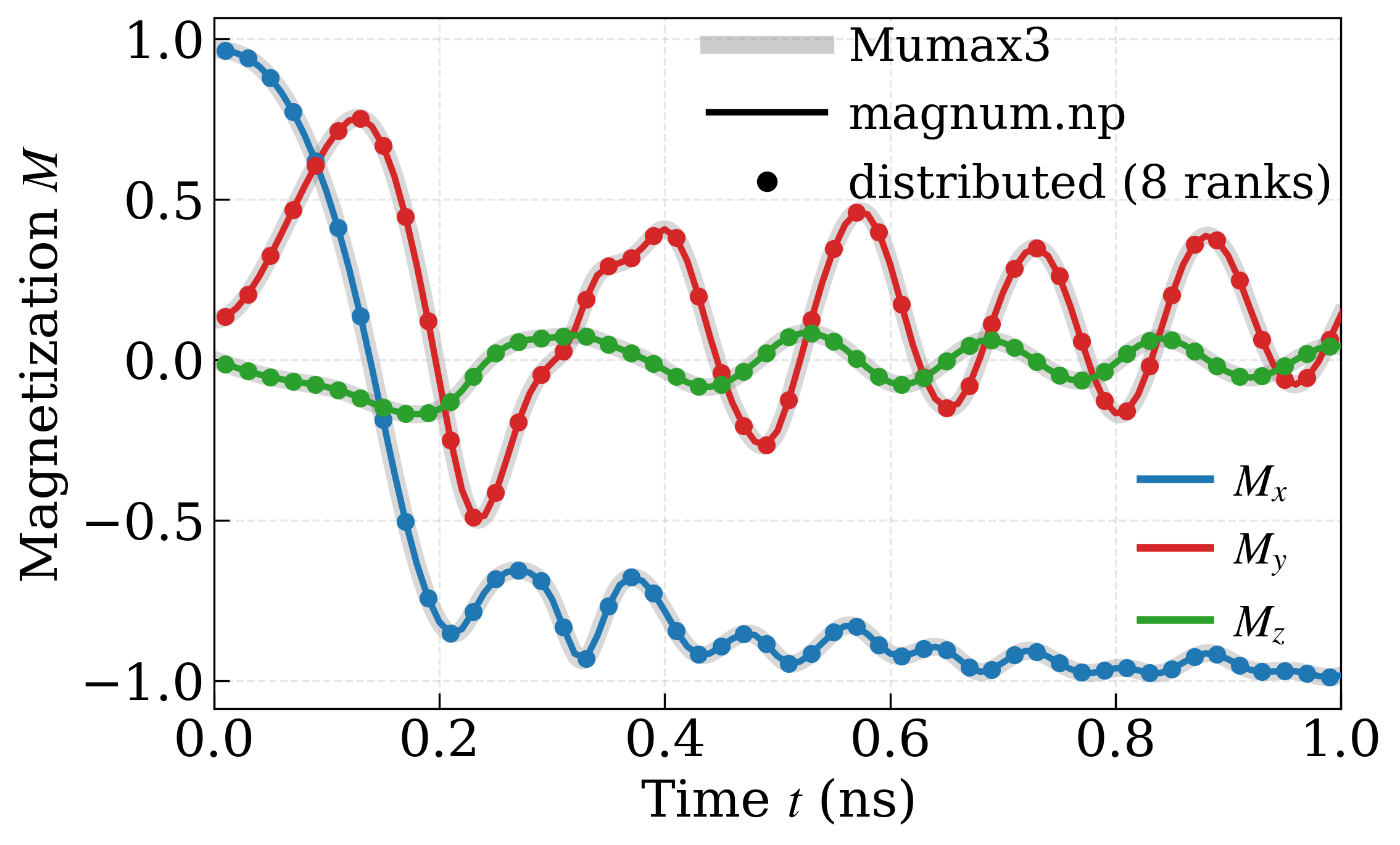}
	\caption{Simulated magnetisation components in the muMAG standard problem 4. The results from Mumax3 and the original non-distributed version of magnum.np were compared with those from the distributed implementation using 8 ranks. The results match both references, validating the implementation of the distributed LLG-RKF45 solver, demagnetisation, and exchange effective field. }
	\label{fig:fig2}
\end{figure}

\subsection*{Performance}
The compute throughput was evaluated on a synthetic cubic mesh with $(N\times N \times N)$ cells, where $N$ ranged from 64K to 64M. All field terms were timed individually over 10,000 iterations, after 1,000 warm-up steps, in double precision. \\

\begin{figure}[ht]
	\centering
	\includegraphics[width=0.99\textwidth]{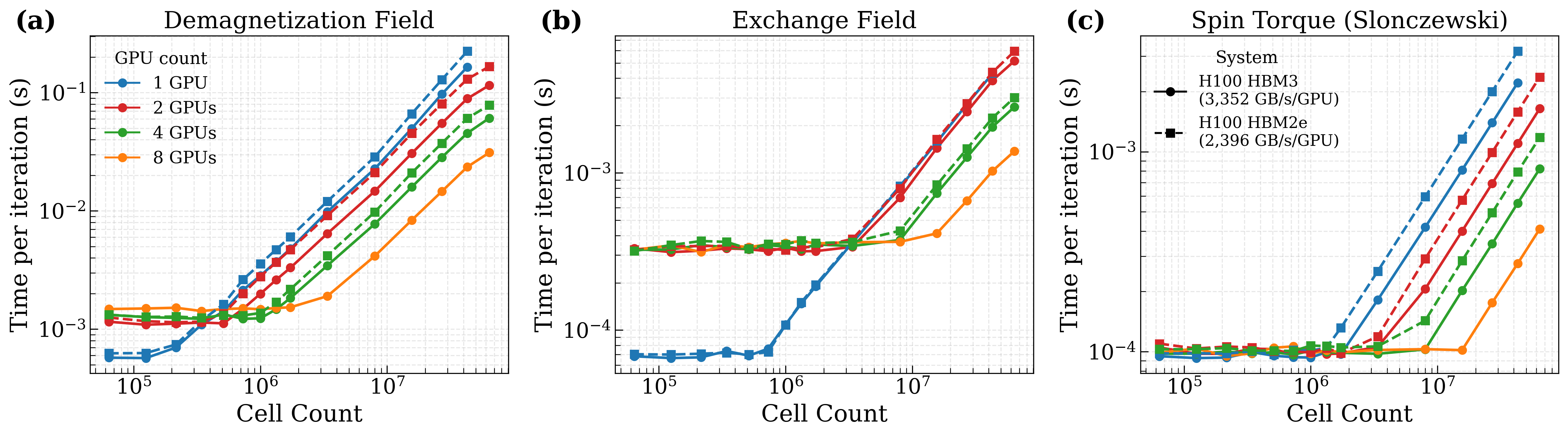}
	\caption{Performance (time per iteration) of various effective fields on two single-node NVIDIA H100 Hopper GPU systems: HBM3 (solid lines, 3,352~GB/s/GPU, up to 8~GPUs via NVSwitch) and HBM2e (dashed lines, 2,396~GB/s/GPU, up to 4~GPUs via NVLink). (a) The demagnetisation field is computationally and communication-intensive; performance gains begin at around $10^6$ cells on both systems. Despite the 40\% difference in memory bandwidth, both systems exhibit near-identical relative scaling, confirming that speedup is governed by the algorithm's communication-to-computation ratio. (b) The exchange field requires a halo exchange at each slab boundary. Significant overhead is observed below $10^6$ cells, and meaningful speedup only appears near $10^7$ cells for 4 and 8 GPUs. (c) Slonczewski spin transfer torque acts only along the z-direction, requiring no inter-rank communication and yielding near-ideal scaling on both systems.}
	\label{fig:fig3}
\end{figure}

First, the multi-GPU performance was benchmarked on two different NVIDIA H100 Hopper GPU systems: one with 8 H100 HBM3 GPUs (bandwidth: 3,352 GB/s/GPU) connected via NVSwitch, and another system with 4 H100 HBM2e GPUs (bandwidth: 2,396 GB/s/GPU) connected via NVLink. The performance evaluation results are shown together in Figure~\ref{fig:fig3} (solid lines: HBM3, dashed lines: HBM2e).  The 40\% difference in memory bandwidth between the two memory generations directly manifests in absolute timing: at N=350 (42.8M cells), the single-GPU HBM2e baseline (224.2~ms) is 1.36$\times$ slower than HBM3 (164.9~ms), consistent with the memory-bandwidth ratio of 1.40$\times$ and confirming that the demagnetisation field is memory-bandwidth bound. Despite this absolute timing difference, both systems exhibit near-identical relative scaling. At N=350, the HBM3 GPUs achieve a near-linear speedup of 7.00$\times$ across 8 GPUs and HBM2e achieves 3.70$\times$ across 4 GPUs, in agreement with HBM3's 3.63$\times$ at the same GPU count. This consistency across memory generations confirms that the scaling behaviour is governed by the algorithm's communication-to-computation ratio rather than raw memory bandwidth. At N=400, the simulation ran out of GPU memory on both systems and could not be completed. For local effective fields such as Slonczewski STT, each rank computes independently, yielding near-ideal speedup on both systems. In contrast, the scaling is poor for effective fields requiring halo exchange. A meaningful speed-up was achieved only beyond 8M cells, reaching 4.2$\times$ with 8 GPUs at N=350 on the HBM3 system. This is attributed to the low compute intensity relative to the demagnetisation field, as well as to CPU command dispatch latency in point-to-point data transfers, as evident in the NVIDIA Nsight Compute analysis. The CPU dispatch overhead can be reduced via a CUDA graph in PyTorch, but that would go against the platform-agnostic design philosophy.\\

A key strength of the PyTorch distributed backend is that scaling to multiple nodes requires zero code changes, only a change to the launcher command. Two compute nodes with 4 NVIDIA H100 HBM2e GPUs (2,396 GB/s/GPU) and 2 Intel Xeon Platinum 8490H 60-core CPUs were connected via an IB switch with two InfiniBand (IB) NDR 400 Gbps Network Interface Cards (NICs), for a total of 8 GPUs. Their performance at N=400 is compared with the single-node HBM3 system in Table~\ref{tab:table1}. It is important to note that this comparison involves two simultaneous differences: a 40\% lower memory bandwidth on HBM2e (2,396 vs.\ 3,352~GB/s/GPU), and a substantially slower inter-GPU interconnect (InfiniBand 400~Gbps vs.\ NVSwitch 900~GB/s bidirectional). Disentangling their contributions is key to interpreting the observed degradation.\\

For local effective fields such as Slonczewski STT that require no inter-rank communication, the single-node benchmarks show that the HBM2e GPU is 1.44$\times$ slower than HBM3 on the same field, consistent with the 1.40$\times$ memory bandwidth ratio. The degradation observed in Table~\ref{tab:table1} for STT is therefore largely attributable to the lower memory bandwidth of HBM2e, with only a modest additional contribution from cross-node barrier synchronization. For halo-exchange fields (Heisenberg exchange, interfacial DMI), the 59\% degradation exceeds what memory bandwidth alone predicts, reflecting the added cost of cross-node point-to-point communication over InfiniBand. The most severe degradation (72\%) is seen for the demagnetisation field, which requires two all-to-all transposes per evaluation: each rank transfers 770~MB of 128-bit complex-number data per collective $(50\times800\times401\times3\times16~\text{bytes})$, translating to 7.7~ms per operation across two 400~Gbps ports, whereas the NVSwitch's 900~GB/s bidirectional bandwidth completes the same transfer in under 1~ms. At this scale, the interconnect bottleneck overwhelmingly dominates, with the memory bandwidth difference playing a secondary role. Although multi-node GPU compute provides no practical benefit at the current problem size, it demonstrates correct and seamless operation and positions the framework to benefit from future high-bandwidth interconnects such as the NVL72 systems from NVIDIA.\\

\begin{table}[ht]
	\caption{Multi-node scaling comparison. Time (ms) per iteration step for different effective fields.}
	\centering
    \begin{tabular}{|l|c|c|}
    \hline
    Effective Field & \begin{tabular}[c]{@{}c@{}}8 H100 HBM3\\ (1 node)\end{tabular} & \begin{tabular}[c]{@{}c@{}}8 H100 HBM2e\\ (2 nodes, 400 Gbps InfiniBand)\end{tabular} \\ \hline
    Demagnetization & 31.2                                                           & 112.4                                                                                 \\ \hline
    Exchange        & 1.4                                                            & 3.4                                                                                   \\ \hline
    Slonczewski STT & 0.4                                                            & 0.6                                                                                   \\ \hline
    \end{tabular}
	\label{tab:table1}
\end{table}

\begin{figure}[ht]
	\centering
	\includegraphics[width=0.99\textwidth]{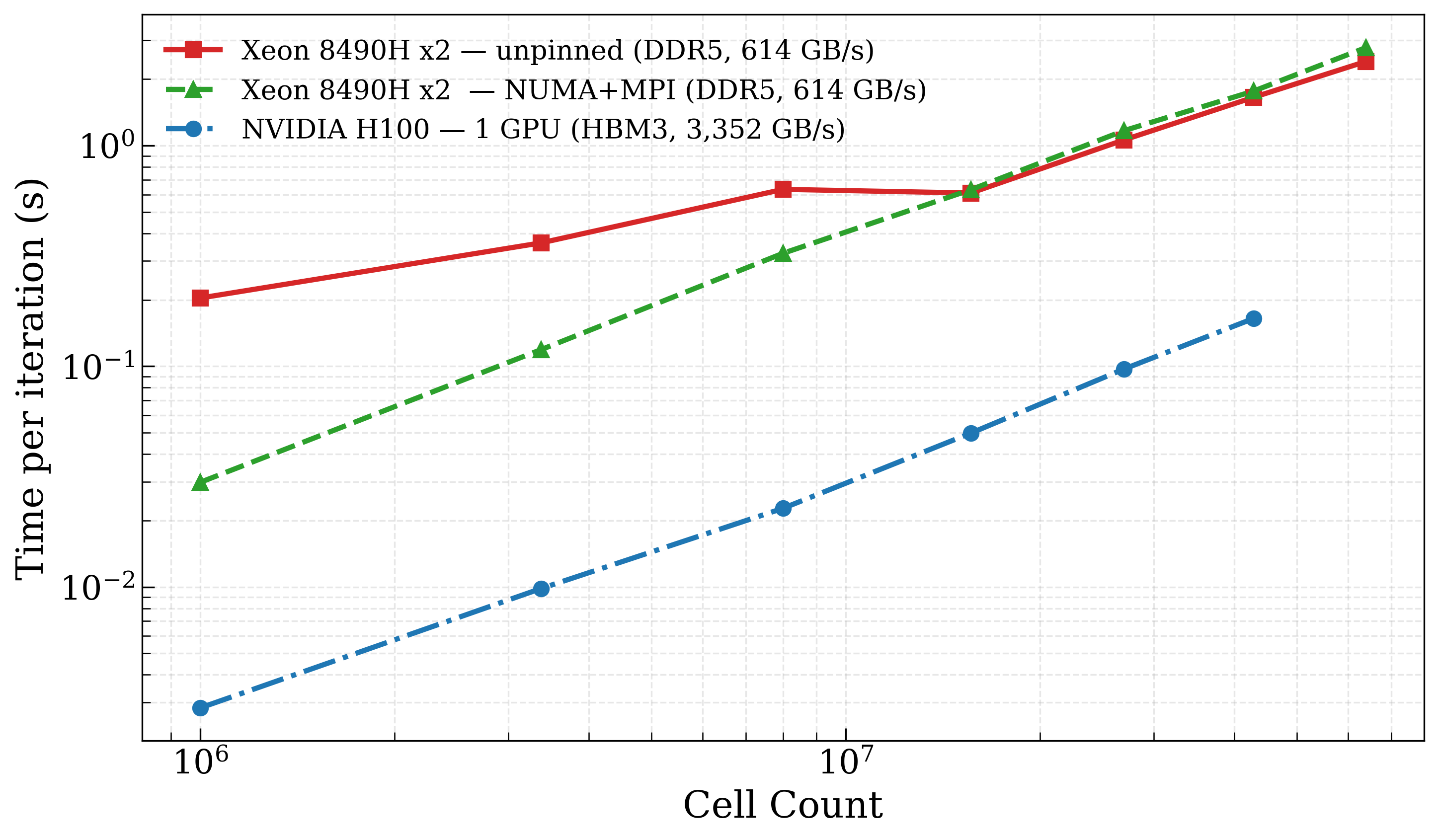}
	\caption{Performance (time per iteration) of the demagnetisation field. Non-distributed magum.np was evaluated on a single NVIDIA H100 HBM3 GPU, and a CPU compute node with 2 Intel Xeon Platinum 8490H 60-core CPUs (no NUMA-pinning). The distributed magnum.np implementation enabled NUMA-based pinning to reduce cross-NUMA communication and yielded a significant performance gain at 106 cells, bringing it to within 10 times slower than a single GPU.}
	\label{fig:fig4}
\end{figure}

We also compiled PyTorch from source using Intel oneAPI 2025 MPI compiler for CPU performance evaluation on a compute node with 2 Intel Xeon Platinum 8490H 60-core CPUs and 614.2 GB/s of total memory bandwidth. Although the 4 compute tiles on each CPU are connected via Embedded Multi-die Interconnect Bridge (EMIB) to behave as a single CPU, each tile has its own memory controller. \cite{sapphire_rapids} Each CPU is thus partitioned into 4 sub-NUMA (non-uniform memory access) clusters for a total of 8 NUMA nodes per compute node in our system. Each MPI rank runs a full parallel process based on OpenMP, forming a hybrid MPI+OpenMP model.
Using the MPI launch command \verb|I_MPI_PIN_DOMAIN=numa|, the simulation performance was compared with and without NUMA pinning, as well as one H100 HBM3 GPU as a reference in Figure~\ref{fig:fig4}. At $10^6$ cells (N=100), representative of widely used $1024 \times 1024$ 2D thin-film problems, NUMA pinning reduces the cross-NUMA traffic and the computation time for the effective demagnetisation field from 204.0 ms to 29.8 ms per step, resulting in a 6.8x speedup. Unfortunately, for larger sizes (N>250), the FFT problem becomes memory-bandwidth-limited regardless of NUMA pinning. 
Even after NUMA pinning, a single H100 GPU remains ~10 to 15 times faster, which is consistent with the hardware specifications. The dual-socket Xeon node provides 614~GB/s of DDR5 bandwidth, approximately 18\% of the H100's 3,352~GB/s HBM3 bandwidth, which directly explains the performance ratio for this memory-bandwidth-limited workload, with additional overhead from cross-socket communication across the UPI interconnect. The CPU can therefore be a viable fallback when GPU availability is tight, even at institutions with GPU clusters. While OOMMF, a widely used CPU-based FEM simulator, supports NUMA-aware compute, it is achieved by allocating memory and compute to a specific NUMA node. This is not suitable for modern many-core CPUs, which are usually partitioned into sub-NUMA nodes with a small number of cores per node (15 cores here). Multinode compute was also tested using a single 200 Gbps IB NIC on socket 0 of the compute node connected via an IB switch. Zero measurable performance gain was observed over single-node, as the all-to-all transpose completely saturates the 200 Gbps port and incurs additional latency when accessing the IB NIC from socket 1.\\

\subsection*{Example - Interfacial-DMI Inducing Multilayer}
Researchers working with interfacial DMI (i-DMI)- inducing layers often use large-scale simulations to fully capture the complex dynamics of maze domains. \cite{cheng2025computational} While it is possible to mathematically concentrate the mesh in the z-direction into one single layer via the effective medium theory \cite{woo2016observation, joos2023tutorial}, the demagnetisation field can dominate in cases of weak i-DMI, leading to a vortex in the domain wall. \cite{legrand2018hybrid} In such cases, it is necessary to simulate the entire stack to account for this effect. We adopted the Pt/Gd/Co/Ni system \cite{10.1063/5.0294523} on a mesh with $(512\times512\times 90)$ cells, each of size $(4\times4\times0.7)$ nm. The system's energy was first minimised, then relaxed to the ground state for more detailed analysis. The final state can be found in Figures 5(a) and (b). At D=0.2 mJ/m$^2$, the demagnetisation field dominates, forming a vortex in the domain wall, as shown in Figure 5(c). With 23.6 million cells, the simulation of the demagnetisation field should benefit from using multiple GPUs. The LLG relaxation process took 8.4 hours to complete on 4 NVIDIA H100 HBM2e GPUs, compared with 50.6 hours on a single GPU. The achieved 6.0x speedup on 4 GPUs surpasses the 3.63x speedup served in the isolated synthetic benchmark at a comparable cell count. This superlinear behaviour is attributed to GPU cache utilisation. 3D FFT is fundamentally memory-bandwidth bound, with the memory controller streaming large, complex buffers. The fully padded FFT for a $(512\times512\times 90)$ mesh exceeds the H100's 50MB L2 cache, leading to memory access from the HBM. When the mesh is distributed across 4 GPUs, each rank's working set is reduced, enabling a larger fraction of the FFT data to reside in the cache. \\

\begin{figure}[ht]
	\centering
	\includegraphics[width=0.99\textwidth]{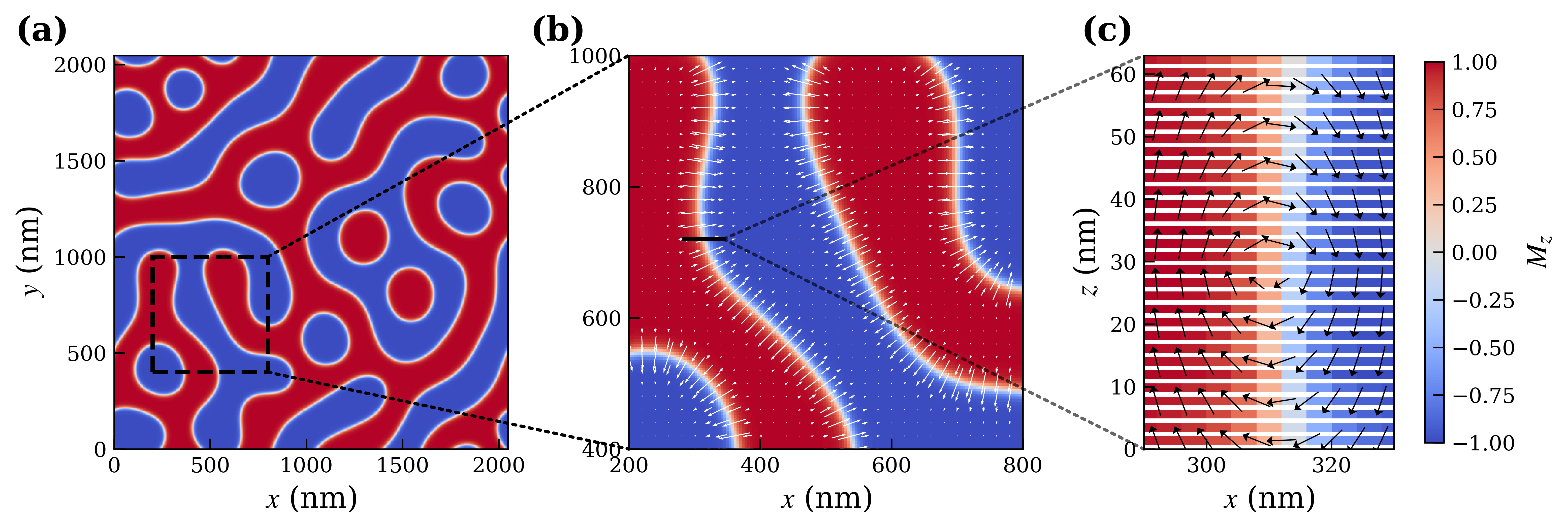}
	\caption{A simulated i-DMI inducing multilayer. (a) Mz at the top layer after relaxation via LLG-RKF45 solver. (b) A zoom-in view of (a) with arrows depicting in-plane components. (c) Cross-section across a domain wall where arrows depict the x and z-components. A vortex is formed in the film with weak i-DMI. }
	\label{fig:fig5}
\end{figure}

\section*{Conclusion}
We present the first Python-native multi-GPU micromagnetic framework. Inheriting the benefits from the non-distributed magnum.np, no C++ or CUDA compilation or knowledge is required for installation or development. The migration from the non-distributed version requires only 8 lines of changes to incorporate a distributed-aware solver, halo exchange, and an all-to-all transpose for the demagnetisation field. A near-linear speedup of 7 was observed for demagnetisation field computation across 8 GPUs connected via NVLink, while halo-exchange scaling is limited by kernel dispatch latency. The flexibility of PyTorch distributed also allows the backend to be changed to MPI for NUMA-pinning on CPU runs. The NUMA-pinned CPU run is 6.8 times faster than the unpinned case, bringing it to within 10x of an H100 GPU for researchers without access to GPUs. Faster simulation enables researchers to explore larger and more complex nanomagnetic systems, accelerating research in magnetic materials and spintronic devices.

\section*{Acknowledgements}
We would like to thank Prof. Terumitsu Tanaka and Prof. Takeshi Nanri for providing value opinion and suggestions.
This work was supported by JSPS KAKENHI (Grant Number 22KK0056, 23K22827, 24H00030, JP24H02235), MEXT Initiative to Establish Next-Generation Novel Integrated Circuit Centers (X-NICS) and The Center for Spintronics Research Network (Osaka). Simulations were performed using the computing resources provided under the General Projects category by the Research Institute for Information Technology, Kyushu University. 

% \section*{Author contributions statement}
% N.A.

\subsection*{Competing interests}
The authors declare no competing interests.

\subsection*{Data availability}
The simulation scripts and data can be found on the code repository \\ \url{https://gitlab.com/jedcheng/magnum-np-distributed}

\bibliography{sample.bib}

\end{document}